# Toward a better understanding of the doping mechanism involved in Mo(tfd-COCF$_3$)$_3$ doped PBDTTT-c

J. Euvrard,[1] A. Revaux,[1] S. S. Nobre,[2] A. Kahn,[3] D. Vuillaume[4,a]

[1]Univ. Grenoble Alpes, CEA-LITEN, Grenoble, 38000, France

[2]Univ. Grenoble Alpes, CEA, Liten, DTNM,SEN,LSIN, F-38000 Grenoble, France

[3]Dept. of Electrical Engineering, Princeton University, Princeton, NJ, 08544, USA

[4]IEMN, CNRS, Univ. Lille, Villeneuve d'Ascq, 59652, France

In this study, we aim to improve our understanding of the doping mechanism involved in the polymer poly[(4,8-bis-(2-ethylhexyloxy)-benzo(1,2-b:4,5-b')dithiophene)-2,6-diyl-alt-(4-(2-ethylhexanoyl)-thieno [3,4-b]thiophene-)-2-6-diyl)] (PBDTTT-c) doped with tris[1-(trifluoroethanoyl)-2-(trifluoromethyl)ethane-1,2-dithiolene] (Mo(tfd-COCF$_3$)$_3$). We follow the evolution of the hole density with dopant concentration to highlight the limits of organic semiconductor doping. To enable the use of doping to enhance the performance of organic electronic devices, doping efficiency must be understood and improved. We report here a study using complementary optical and electrical characterization techniques, which sheds some light on the origin of this limited doping efficiency at high dopant concentration. Two doping mechanisms are considered, the direct charge transfer (DCT) and the charge transfer complex (CTC). We discuss the validity of the model involved as well as its impact on the doping efficiency.

**I. INTRODUCTION**

Controlled p and n-doping has been a key factor in the success of inorganic semiconductor devices.[1] The ability to dope a semiconductor leads to improved transport and interface properties. Organic semiconductors are usually undoped and exhibit very small conductivity, and doping is essential to reduce ohmic losses and obtain efficient contacts.[2] Controlled doping can lower the electric field required to drive organic light emitting diodes,[3,4] avoid voltage drops at the active layer-electrode interface of organic solar cells [5,6] or reduce the contact resistance in organic field effect transistors.[7–9] The development of efficient, controllable and stable n and p-dopants remains therefore an important challenge..

---

[a] Electronic mail: Dominique.vuillaume@iemn.fr

A recurring issue with doping in organic semiconductors is the need for high dopant concentrations, which leads to the degradation of transport properties.[10,11] A better understanding of the mechanisms involved in doping processes is therefore of great importance. Although the limited doping efficiency at low doping concentration has been attributed to the filling of trap states in the polymer host[12,13], the origin of low doping efficiency at high doping concentration has not been clearly established yet, and its elucidation could provide clues for dopant design improvement.

In organic semiconductors, doping is achieved by adding appropriate atoms or molecules to the host matrix. In the case of p-doping, the process is usually described in terms of a direct transfer of an electron from the HOMO of the host to the LUMO of the dopant, as illustrated in Fig. 1 (a).[14,15] This doping mechanism, called direct charge transfer (DCT), has been identified in multiple molecular and polymer semiconductor-dopant mixtures with the formation of sub-bandgap absorption peaks associated with the dopant ion or polarons in the host.[16–18]

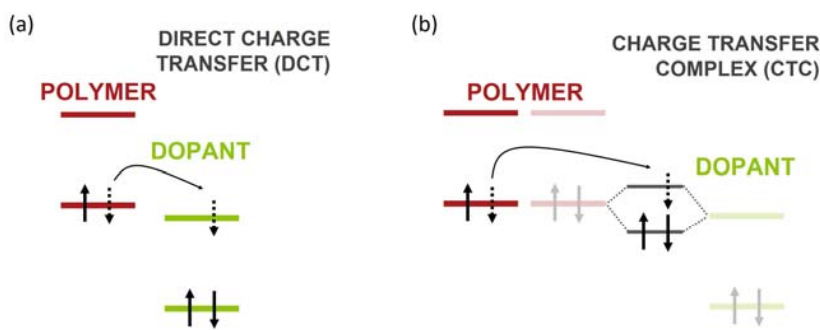

FIG. 1. Schematic of the direct charge transfer (DCT) (a) and Complex Charge Transfer (CTC) (b) models used to describe the doping process in organic semiconductors, here in the case of p-doping.

However, some studies carried out on organic host-dopant mixtures can raise questions regarding the DCT hypothesis. For MeO-TPD doped with $C_{60}F_{36}$ and $F_6TCNNQ$, Tietze et al.[19] have observed a Fermi level pinning a few hundred meV above the polymer HOMO.[14] For some polymer-dopant blends, new occupied states have been observed in the semiconductor bandgap.[20–22] It has also been shown that p-dopants with a LUMO level lying above the polymer HOMO can lead to effective p-doping.[19,23] It is still unclear whether such observations could be explained with DCT hypothesis or whether it suggests the involvement of an alternative doping mechanism. An alternative model involving the formation of a charge transfer complex (CTC) has been proposed by Salzmann et al.[14,24].

The CTC model is illustrated in Fig. 1 (b) for p-doping. The orbital overlap between the semiconductor HOMO and the dopant LUMO can lead to an energy level splitting with formation of supramolecular hybrid orbitals (SMHO), the bonding and antibonding states.[24] To p-dope the semiconductor, the electron must be transferred from the semiconductor HOMO to the antibonding state of the complex.[11] An uphill charge transfer is then attributed to this process, which would lead to lower doping efficiencies.

Density Functional Theory (DFT) calculations have been carried out on various organic semiconductor-dopant mixtures, highlighting the formation of a CTC with an antibonding level lying several hundred meV above the semiconductor HOMO for p-type doping.[24–26] These antibonding states have also been measured by inverse photoemission spectroscopy (IPES) for F$_4$TCNQ doped 4T at a high doping concentration (1:1), although this mixture is an alloy rather than a doped semiconductor.[17] A doping acceptor level situated a few hundreds of meV above the semiconductor HOMO can explain the phenomenon of Fermi level pinning as the probability of dopant ionization is reduced when the Fermi level crosses the antibonding state.[27]

Although a few studies have been conducted to determine what model is involved in different organic semiconductor-dopant mixtures, the underlying origins and consequences of each doping mechanism remain to be understood.[14] In this study, we aim to determine what model (DCT or CTC) corresponds to our polymer-dopant mixture, and to deepen our understanding of the mechanism involved through complementary optical and electrical characterization techniques. We also discuss the limits and open questions regarding these models as well as the consequences on the electrical performances of the doped layer.

This study is carried out on the polymer Poly[(4,8-bis-(2-ethylhexyloxy)-benzo(1,2-b:4,5-b')dithiophene)-2,6-diyl-alt-(4-(2-ethylhexanoyl)-thieno [3,4-b]thiophene-)-2-6-diyl)] (PBDTTT-c) p-doped with molybdenum tris[1-(trifluoroethanoyl)-2-(trifluoromethyl) ethane-1,2-dithiolene] (Mo(tfd-COCF$_3$)$_3$), a soluble derivative of Mo(tfd)$_3$. The molecular structures of both polymer and dopant are shown in Fig. 2. In a previous study,[28] we have provided evidence of the effective p-doping of PBDTTT-c with this p-dopant and highlighted the limited doping efficiency at high doping concentration.

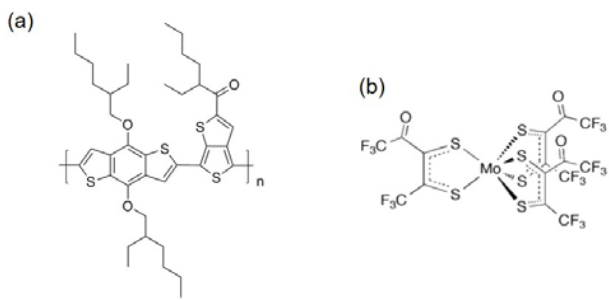

FIG. 2. Molecular structure of Poly[(4,8-bis-(2-ethylhexyloxy)-benzo(1,2-b:4,5-b')dithiophene)-2,6-diyl-alt-(4-(2-ethylhexanoyl)-thieno [3,4-b]thiophene-)-2-6-diyl)] (PBDTTT-c) (a) and molybdenum tris(1-(trifluoroacetyl)-2-(trifluoromethyl)ethane-1,2-dithiolen (b).

## II. EXPERIMENTAL METHODOLOGY

Schottky diodes are processed for capacitance and admittance spectroscopy measurements. Poly(3,4-ethylenedioxythiophene)-poly(styrenesulfonate) (PEDOT:PSS) Orgacon HIL 1005 from Agfa is used as ohmic bottom contact and aluminum as Schottky top contact. To avoid any interface effects, we sandwich the layer of interest with thin layers of pure polymer, leading to the following structure: glass/ITO/PEDOT:PSS/PBDTTT-c/pure or doped PBDTTTc/PBDTTT-c/Aluminum. The 110 nm thick ITO is patterned by photolithography on cleaned ITO covered glass substrates. PEDOT:PSS is spin-coated on UV-ozone treated ITO to reach a thickness of 150 nm and annealed at 115°C under nitrogen for 10 min. A layer of PBDTTT-c (70±10 nm) is then spin-coated on PEDOT:PSS and annealed at 115°C under nitrogen for 10 min as well. The layer of interest, pure or doped at 0.5, 1, 2 and 5% molar ratio (MR), is deposited on a cleaned silicon substrate annealed at 115°C under nitrogen for 10 min and laminated following the soft contact transfer lamination (SCTL) process described elsewhere.[29] Another 70 nm thick layer of pure PBDTTT-c is laminated on top of the structure. Following this double lamination step, a 100 nm thick aluminum top electrode is evaporated through a shadow mask in a vacuum chamber. To avoid unintentional doping due to oxygen, the diodes are encapsulated in glovebox (with less than 3 ppm of $H_2O$ and $O_2$) with glass using an epoxy glue.

Layers of PBDTTT-c, intrinsic and p-doped with $Mo(tfd-COCF_3)_3$, are processed on glass substrates to carry out UV-visible and photoluminescence spectroscopy measurements. Eight different molar concentrations are processed from 0.5% to 6% MR to cover the different regimes observed through electrical characterization. The concentrations of PBDTTT-c and $Mo(tfd-COCF_3)_3$ have been carefully chosen to obtain similar thicknesses. The solutions are spin-coated on borosilicate glass substrates and annealed at 115°C for 10 min under nitrogen. The film thicknesses are measured with a contact profilometer and the average thickness for all layers is estimated at 390±10 nm.

To extract the hole density, capacitance-voltage C(V) measurements are carried out in the dark using a probe-station and an LCR meter Agilent E4980A. The capacitance is measured for a DC bias varying from -5 V to +2 V with an AC signal of 100 mV in amplitude and a frequency of 100 Hz.

The absorption spectra are obtained from reflection and transmission measurements using a LAMBDA 950 UV/Vis/NIR Spectrophotometer from Perkin Elmer with an integrating sphere.

Photoluminescence measurements are carried out with a modular Fluorolog FL 3-22 spectrofluorimeter from Horiba-Jobin Yvon-Spex with a near infra-red photomultiplier from Hamamatsu (T5509-73). The size of the slits and orientation of the sample are optimized with a sample of pure polymer and kept constant for all measurements.

Time resolved photoluminescence (TRPL) measurements are performed with a titanium-sapphire laser in pulse mode. The laser emits pulses at a wavelength of 810 nm with a duration of 200 fs. The laser has been set, with a cavity dumper, to emit pulses at a rate of 1 MHz. We use a nonlinear crystal to generate the second harmonic, to get an excitation wavelength of 405 nm. The laser average power was set at 65 nW for all samples; that corresponds to 65 fJ per pulse. A monochromator is used to select the emission wavelength, and a silicon avalanche photodiode (APD) measures the photoluminescence signal, with a time resolution of 300 ps.

For admittance spectroscopy analysis, capacitance and conductance measurements are carried out at temperatures ranging from 100 to 350 K. The measurements are performed in a vacuum chamber with a chuck cooled down to 77 K with liquid nitrogen and heated to reach the required temperature. An impedance analyzer Keysight E4990A is used. Capacitance C(f) and conductance G(f) characteristics are carried out from 20 Hz to 10 MHz with an AC signal amplitude of 10 mV and a bias of 0 V.

### III. RESULTS AND DISCUSSION
#### A. Hole density with dopant concentration

The addition of dopant molecules in the polymer matrix leads to the transfer of holes toward the polymer HOMO. To quantify the creation of holes in the semiconductor, we measure the hole density $p$ as a function of dopant concentration in the polymer. This measurement is achieved through Mott-Schottky analysis, with the hole density extracted from C(V) measurements according to the following equation:[30]

$$N_A^- = -\frac{2}{q\varepsilon_0\varepsilon_r A^2 \frac{d\left(\frac{1}{C^2}\right)}{dV}}, \quad (1)$$

where $N_A^-$ is the density of ionized dopant molecules ($N_A^- = p$ if all ionized dopants lead to free carriers), q is the elementary charge, $\varepsilon_0$ the vacuum permittivity, $\varepsilon_r$ the relative permittivity of the semiconductor and A the area of the diode. The density of ionized dopant molecules and therefore the hole density is extracted for the pure polymer and four doping concentrations ranging from 0.5% to 5% MR. For pure PBDTTT-c, we obtain a hole density around $5.10^{15}\ cm^{-3}$. However, this value might be overestimated due to the limit of the Mott-Schottky analysis at low carrier densities as explained by Kirchartz et al.[31] and detailed in supplementary material (Fig. S1). Fig. 3 shows the evolution of the hole density as a function of doping concentration. The C(V) measurements and the Mott-Schottky plots are given in supplementary material Fig. S2 along with a detailed explanation of the hole density extraction.

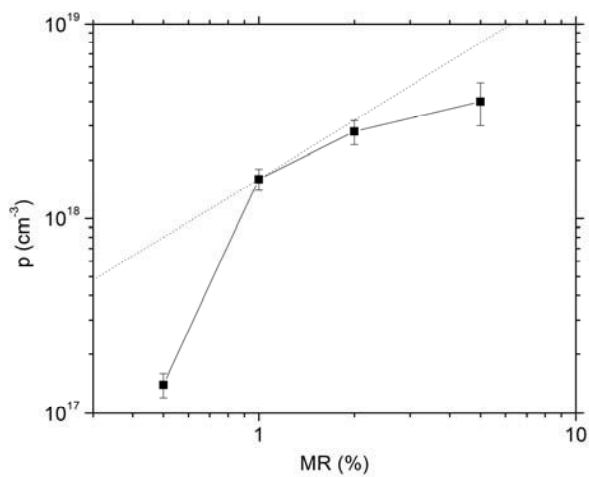

FIG. 3. Hole density as a function of dopant concentration in molar ratio (MR). A linear function is given in dotted line.

The hole density exhibits a superlinear increase up to a concentration of 1% MR and a sublinear increase above this threshold. The superlinear increase at low doping concentration is likely related to trap filling, as already identified for several other doped organic semiconductors.[2,32,33] The origin of the sublinear increase at high dopant concentration remains to be understood. One of the factors could be dopant aggregation, which limits the interaction with the polymer host.[34,35] However, in a previous study, we highlighted the formation of aggregates above a concentration of 2% MR.[28] Nuclear Magnetic Resonance measurements indicated that all dopant molecules added to the matrix react with the polymer. Therefore, no aggregates of pure dopant are formed. Although the formation of polymer-dopant aggregates could contribute to the degradation of the transport properties above 2% MR, only the formation of pure dopant phases could explain the sublinear increase of the hole density at high dopant concentration through a reduction of charge transfer from the polymer HOMO toward the dopant LUMO.

As an alternative hypothesis, we can consider acceptor levels situated a few hundred meV above the HOMO and, therefore, possessing a lower probability of ionization. Above a certain dopant concentration threshold, new additional dopant molecules are not ionized, leading to the saturation of the Fermi level around the acceptor level. In organic semiconductors, the phenomenon of Fermi level pinning has been observed for several polymer-dopant mixtures, with the saturation of the Fermi level a few hundred meV above the polymer HOMO.[17,36,37] Tietze et al.[19] have highlighted the relation between the doping efficiency decrease and the position of the acceptor state in the bandgap. This observation would be consistent with the formation of a CTC between the polymer and the dopant, with an antibonding state lying several hundred meV above the polymer HOMO. Temperature-dependent C(V) measurements could be used to analyze the activation energy of doping and, therefore, identify the position of the acceptor state above the polymer HOMO. Unfortunately, the small temperature range permitted by our samples limits the extraction of the doping activation energy as explained in supplementary material (Fig. S4.). To further understand the evolution of the hole density at high dopant concentration, we need to identify the doping mechanism involved in our polymer-dopant mixture.

**B. Sub-bandgap absorption**

It has been demonstrated that both DCT and CTC result in the modification of the polymer absorption spectrum in the sub-bandgap region.[38] In case of DCT, the sub-bandgap absorption peaks originate either from the ionized dopant or from the polarons in the polymer.[16] When a CTC is formed between the polymer and the dopant, a transition is expected from the bonding to the antibonding level of the complex. As the gap of the CTC is usually lower than the polymer bandgap, an absorption peak can be observed in the sub-bandgap region.[17]

UV-visible absorption spectroscopy is therefore carried out on Mo(tfd-COCF$_3$)$_3$ doped PBDTTT-c with the doping concentration varying from 0% to 6% MR to observe the impact of the polymer-dopant interaction on the sub-bandgap absorption. The absorption coefficient $\alpha = 4\pi k/\lambda$ is calculated for each sample through the determination of the extinction coefficient k using the OptiChar module of Opti-Layer Thin Film Software as a function of the wavelength λ of the incident photons. Fig. 4 shows the absorption coefficient spectra for each doping concentration in the sub-bandgap region. We observe the formation of two absorption peaks at 860 meV and 1.1 eV upon addition of dopant, and their intensity increases with the doping concentration. A third peak may also appear upon doping below 600 meV. Absorption measurements in the far infra-red would be necessary to validate this observation and determine the position of the peak. Such peak could be due to the formation of polarons in the polymer, as determined for P3HT doped with F$_4$TCNQ.[17]

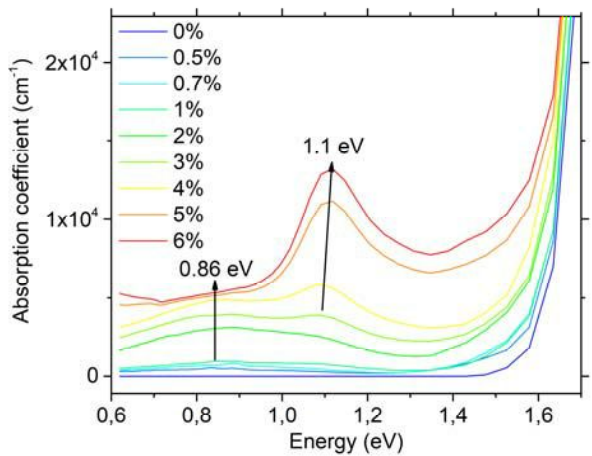

FIG. 4. Absorption coefficient spectra for the pure polymer and 8 doping concentrations from 0.5% to 6% MR.

As both DCT and CTC models result in the formation of sub-bandgap absorption peaks, we need to study both possibilities. If DCT is involved in the dopant interaction with PBDTTT-c, the sub-gap absorption would be due to the dopant in its ionized form or to polarons in the polymer. However, Mohapatra et al.[39] have measured the absorption spectra of $Mo(tfd-COCF_3)_3$ in neutral, monoanionic and dianionic oxidation states, and none of these corresponds to the absorption peaks created in the doped polymer. The anionic forms of the dopant exhibit strong absorption peaks below 800 nm and therefore within the absorption spectrum of the polymer. A small broad peak is visible around 950 nm for the monoanionic oxidation state of the dopant but its intensity might be too low to be observed on the doped polymer spectra. Yet, DCT cannot be ruled out at this stage as polarons in the polymer could be responsible for the sub-bandgap absorption peaks.

Considering now the formation of a CTC between the polymer HOMO and the dopant LUMO, we can determine the position of the corresponding bonding and antibonding levels. According to Méndez et al.[26], the magnitude of the energy-level splitting is described by a Hückel-like model. The gap of the complex depends on the polymer ionization energy $IE_P$ and the dopant electron affinity $EA_D$, but also on the intermolecular electronic coupling described by the resonance integral $\beta$:[17]

$$E_{gap}^{CTC} = 2\sqrt{(IE_P - EA_D)^2 + 4\beta^2} \,. \tag{2}$$

The ionization energy of PBDTTT-c has been measured by ultraviolet photoelectron spectroscopy (UPS) around 5.15 eV.[40] The determination of the dopant electron affinity is not straightforward. Because of difficulties in the formation of thin films of the dopant, its electron affinity has not been measured directly by inverse photoemission spectroscopy (IPES). An indirect measurement using cyclic voltammetry gives a value around 5.65 eV.[41] However, the uncertainty associated with cyclic voltammetry is significant (error margins larger than 0.1 eV)[42] and Sworakowski suggests that gaps measured by electrochemical techniques are smaller than transport gaps.[43] Moreover, recent experiments suggest a value closer to 5.3 eV.[44] We also need to consider the uncertainty of the ionization energy and electron affinity of the two materials when they are mixed. Therefore, we consider a dopant electron affinity $EA_D$ between 5.3 and 5.65 eV. Taking into account a gap $E_{gap}^{CTC}$ of 1.1 eV, corresponding to the position of the main peak, we obtain $\beta$ between 0.11 and 0.26 eV.

In a Hückel-like model, the values of the complex bonding $E_{CTC}^B$ and anti-bonding $E_{CTC}^A$ levels are given by the following equation:[38]

$$E_{CTC}^{B/A} = \frac{IP_P + EA_D}{2} \pm \sqrt{(IE_P - EA_D)^2 + 4\beta^2}. \qquad (3)$$

Considering $\beta$ between 0.11 and 0.26 eV, the antibonding level of the complex would be situated between 300 and 470 meV above the polymer HOMO as illustrated in green in Fig. 5.

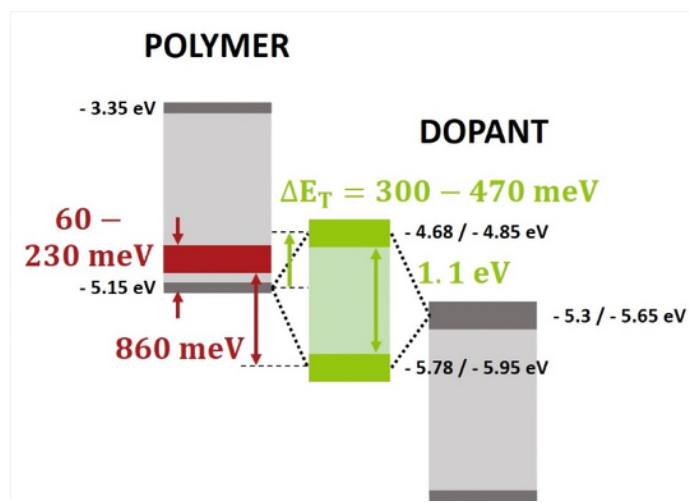

FIG. 5. Schematic of the potential CTC formed between the PBDTTT-c HOMO and the Mo(tfd-COCF$_3$)$_3$ LUMO with the energy levels obtained from the Hückel-like model and intrinsic trap level suggested in the polymer bandgap.

The origin of the second peak centered around 860 meV is also related to the doping process as this peak is not visible for the pure polymer. However, UV-visible absorption measurements carried out on different batches of polymer doped with the same dopant and at the same concentration (Fig. S5 in supplementary material) reveal that the presence of this peak depends on the batch of polymer used and, therefore, might be due to a defect or impurity in the polymer. Considering the CTC model, a transition from the bonding level of the complex toward a trap state in the polymer bandgap could result in the formation of an additional sub-bandgap peak at 860 meV. The corresponding energy level would be situated between 60 and 230 meV above the polymer HOMO, as illustrated in red in Fig. 5.

At this stage of the study, we highlighted the formation of sub-bandgap absorption peaks, which could be due to the formation of polarons in the polymer upon DCT or to the formation of a CTC between the polymer HOMO and the dopant LUMO. Considering the hypothesis of CTC, we extracted the corresponding energy levels lying in the polymer bandgap within an accuracy range limited by the knowledge of the dopant electron affinity.

**C. Fluorescence quenching**

Determining the evolution of the polymer photoluminescence (PL) intensity can further help identify the doping mechanism. Fig. 6 shows the PL spectra of pure and doped PBDTTT-c highlighting a strong quenching of the fluorescence upon addition of dopant molecules.

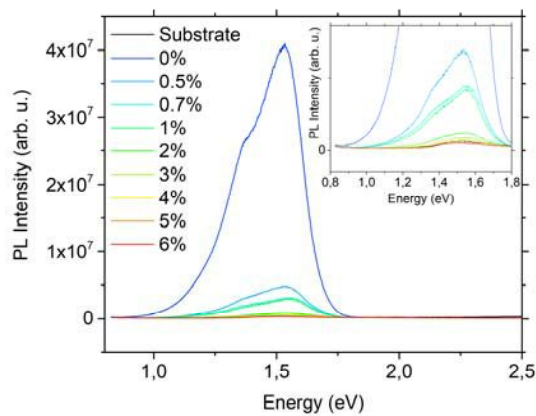

FIG. 6. Photoluminescence (PL) emission spectra for pure and doped PBDTTT-c and for the borosilicate glass substrate at an excitation wavelength of 350 nm. Magnified view of the PL spectra for doped PBDTTT-c given as inset.

If we consider that hole polarons are formed upon dopant addition in the polymer matrix through DCT mechanism, we would expect to observe fluorescence quenching with increasing doping concentration. The absorption of a photon by the polymer leads to the formation of an exciton, which can diffuse over a path of approximately 10 nm.[45,46] The creation of a polaron by ionization of the polymer leads to the formation of two localized states inside the bandgap.[14] If the exciton reaches a hole polaron, a non-radiative path is allowed and competes with the radiative recombination observed in the pure polymer. This process is called dynamic quenching. Yu *et al.*[47] have observed fluorescence quenching in MEH-PPV due to polarons created by charge injection in the layer.

The formation of a CTC between the polymer HOMO and dopant LUMO can induce fluorescence quenching as well. We have highlighted in Fig. 3 the sublinear increase of the hole density at high doping concentration. As a consequence, we expect a significant amount of antibonding states to be empty as they don't participate to the p-doping. The availability of an energy level in the bandgap can also induce dynamic quenching. Moreover, the presence of nonradiative complexes contributes to the photoluminescence decrease, as some photons are directly absorbed by the complex itself. This process is called static quenching. Tyagi *et al.*[48] have shown that the photoluminescence of $Alq_3$ is quenched when the dopant $F_4TCNQ$ is introduced. This observation confirmed their hypothesis of CTC formation formulated according to UV-visible absorption spectroscopy measurements.

Determining the type of quenching involved, static or dynamic, will therefore provide indications on the doping mechanism. If the fluorescence quenching of PBDTTT-c with $Mo(tfd-COCF_3)_3$ possesses a static component, it strengthens the hypothesis of CTC formation between both components. If the quenching is purely dynamic, the hypothesis of polarons, and therefore DCT, is more likely.

Static and dynamic quenching are described by the Stern-Volmer equation:[49]

$$\frac{I_0}{I} = 1 + K_{SV}[Q], \tag{4}$$

where $I_0$ and $I$ are the emission intensities in the absence and presence of quencher (dopant) respectively, $K_{SV}$ is the Stern-Volmer constant and $[Q]$ the molar concentration of the quencher. The emission intensities of pure and doped PBDTTT-c have been extracted by fitting the spectra with Gaussian distributions and considering the area under the curve. The baseline is subtracted before fitting as the signal to noise ratio is low at high doping concentration. Fig. 7 (a) shows the Stern-Volmer plot with $[Q]$ determined with respect to the concentration of fluorophore (PBDTTT-c). Although we expect a linear dependence for static or dynamic quenching as suggested by equation 4, a deviation from the linear increase is observed for doping concentrations above 2% MR. A positive deviation can be explained by a combination of static and dynamic quenching, or by a large extent of quenching called sphere of action.[49]

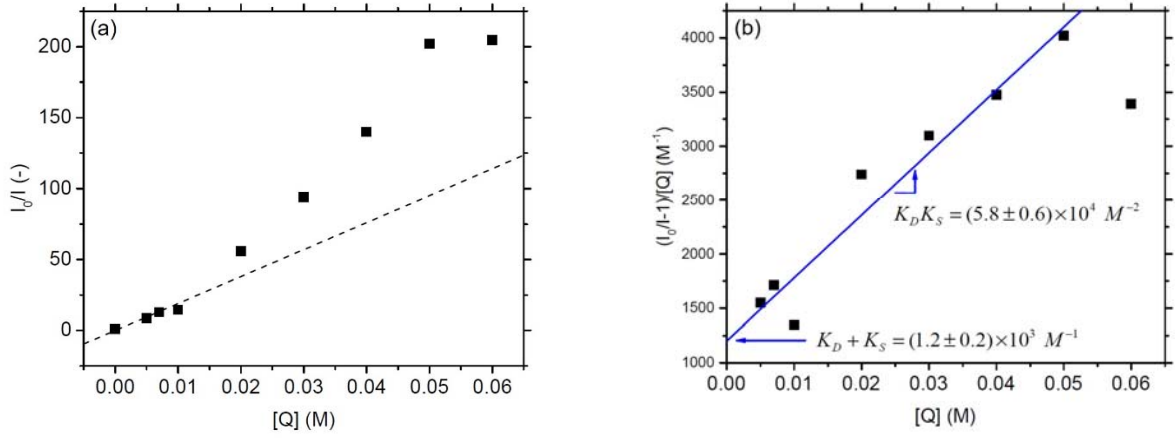

FIG. 7. Stern-Volmer plot determined with the total area under the curve and linear function (dotted line) (a). Corresponding Stern-Volmer plot to separate $K_D$ and $K_S$ (b).

When the presence of quenchers involves both static and dynamic quenching of the fluorophore, the Stern-Volmer equation is modified as follows:[49]

$$\frac{I_0}{I} = (1 + K_D[Q])(1 + K_S[Q]). \tag{5}$$

The dynamic constant $K_D$ can be determined using time resolved photoluminescence (TRPL) measurements and the constant for complex formation $K_S$ can be deduced from a modified representation of the Stern-Volmer plot. Since equation 5 can be rewritten

$$\frac{\left(\frac{I_0}{I}-1\right)}{[Q]} = (K_D + K_S) + K_D K_S[Q], \tag{6}$$

the plot of $(I_0/I - 1)/[Q]$ with respect to the concentration of quencher $[Q]$ leads to a straight line with slope $K_D K_S$ and intercept $K_D + K_S$. Fig. 7 (b) shows the modified representation of the Stern-Volmer plot, which exhibits a linear increase indicating that both dynamic and static mechanisms might be involved in the quenching of PBDTTT-c.

To verify the presence of dynamic quenching and determine $K_D$, we analyze the evolution of the fluorophore excited state lifetime $\tau$ with doping. $\tau$ depends on the radiative $\tau_R$ and non-radiative $\tau_{NR}$ lifetimes ($1/\tau = 1/\tau_R + 1/\tau_{NR}$). In case of dynamic quenching, the addition of quenchers offers non-radiative pathways to the excited fluorophore decreasing $\tau_{NR}$ and, therefore, the fluorophore excited state lifetime $\tau$. We can show that $\tau$ follows the evolution of the fluorescence intensity $I$:[49]

$$\frac{\tau_0}{\tau} = \frac{I_0}{I}, \tag{7}$$

with $\tau_0$ the fluorophore excited state lifetime when no quencher is added. However, fluorescence quenching due to the formation of non-radiative complex does not affect any of the radiative or non-radiative lifetimes of the fluorophore. Therefore, the ratio $\tau_0/\tau$ is unity for static quenching.

Fig. 8 (a) shows the TRPL spectra for pure and doped PBDTTT-c with concentrations ranging from 0.5% to 4% MR. A double exponential fit with $\tau_1$ and $\tau_2$ is used to extract the excited state lifetime. The time constant $\tau_1$ is due to the APD response. The time constant $\tau_2$ corresponds to the excited state in the polymer and needs to be deconvoluted from the APD photodetector response time as detailed in supplementary material. The PBDTTT-c excited state lifetime decreases from 780±10 ps without quencher to 350±20 ps with a doping concentration of 4% MR. The ratio $\tau_0/\tau$ with respect to the concentration of quencher [Q] is given in Fig. 8 (b) and exhibits a linear increase, as expected if dynamic quenching is involved. We extract a dynamic constant $K_D$ of 36±2 M$^{-1}$. The bimolecular quenching constant $k_q$ is defined by $K_D = k_q \tau_0$, leading to a value of $(4.6 \pm 0.3) \times 10^{10}\ M^{-1}s^{-1}$, which is consistent for dynamic quenching due to diffusion.[49] Using the modified version of the Stern-Volmer plot (Fig. 7 (b)), we can determine the constant for complex formation $K_S$ of $(1.2 \pm 0.3) \times 10^3\ M^{-1}$. This value is of the same order of magnitude as MEH-PPV complexed with $C_{60}$ with a constant of $2.5 \times 10^3\ M^{-1}$.[50] However, $K_S$ is 2 orders of magnitude higher for Mo(tfd-COCF$_3$)$_3$ doped PBDTTT-c than for F$_4$TCNQ doped Alq$_3$ with a constant of $13.8\ M^{-1}$.[48] Given that the electron affinity of F$_4$TCNQ (5.2 eV) is lower than the ionization potential of Alq$_3$ (5.7 eV), this observation is consistent with a lower probability for complex formation between these components.

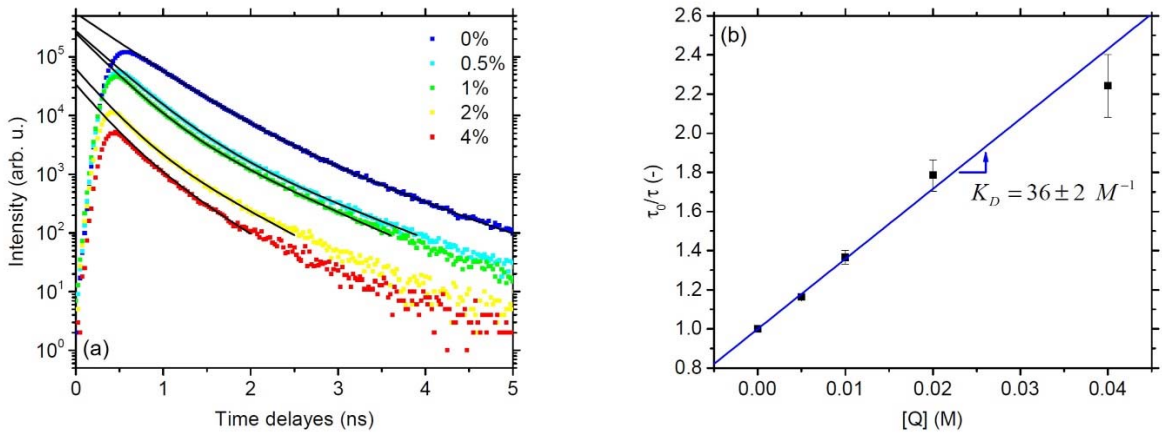

FIG. 8. TRPL spectra for 0%, 0.5%, 1%, 2% and 4% MR doped PBDTTT-c and double exponential fit (black lines) in a semi-logarithmic scale (a). Plot of $\tau_0/\tau$ with respect to the doping concentration and extraction of dynamic constant $K_D$ (b).

Therefore, using photoluminescence spectroscopy and TRPL measurements, we have highlighted the fluorescence quenching of PBDTTT-c upon addition of dopant molecules. $K_D$ and $K_S$ have been extracted and are consistent with the literature. The involvement of both static and dynamic quenching strengthens the hypothesis of CTC.

**D. Gap states formation**

To validate the assumption of CTC formation suggested by photoluminescence measurements, we need to probe the existing energy levels in the polymer bandgap and compare their position with the results obtained from UV-visible spectroscopy considering the CTC hypothesis (Fig. 5). Admittance spectroscopy performed on Schottky diodes is used to probe the energy levels in pure and doped PBDTTT-c. This technique consists in analyzing the capacitance C or conductance G with frequency and at different temperatures. Upon addition of an oscillating signal, the trapping and de-trapping of carriers by the energy levels in the band gap (trap states) impacts both capacitance and conductance values. This response is temperature dependent and enables the extraction of parameters related to the traps. To observe the signature of trap states, we plot $(G - G_0)/\omega$ with respect the angular frequency $\omega$, where $G_0$ is the value of the conductance at the lowest frequency measured. A trap state in the polymer bandgap leads to the formation of a peak centered at $\omega_T$ in the conductance spectrum and this peak shifts with temperature following an Arrhenius law:[51]

$$\omega_T = \nu_0 exp\left(-\frac{E_T}{k_B T}\right) , \qquad (8)$$

where $\nu_0$ corresponds to the attempt-to-escape frequency, $E_T$ the activation energy of the trap state, $k_B$ the Boltzmann constant and T the temperature.

Fig. 9 shows the conductance spectra for the pure polymer (a) and for two doping concentrations, 2% (c) and 5% MR (e). The corresponding C(ω) plots are given in supplementary material (Fig. S6). For pure PBDTTT-c, a peak is visible for temperatures above 300 K, suggesting the presence of an energy level in the polymer bandgap. To determine the activation energy associated with this energy level, we plot $\ln(\omega_T)$ as a function of $1/k_B T$ in Fig. 9 (b). For pure PBDTTT-c, an activation energy around 200 meV is obtained. However, precautions need to be taken on this value due to the low conductance signal. The existence of trap states in the polymer host is in line with the phenomenon of trap filling identified in Fig. 3. When the polymer is doped at 2% MR, two peaks can be observed in the conductance spectra. In order to properly determine $\omega_T$ for both peaks with temperature, the spectra are fitted with Gaussian functions as shown in Fig. S7 in supplementary material. Activation energies of $280 \pm 20\ meV$ and $480 \pm 20\ meV$ are extracted at this doping concentration in Fig. 9 (d). Increasing the doping concentration from 2% to 5% MR does not lead to major changes in the position of the trap states. At 5% MR, activation energies of $220 \pm 10\ meV$ and $430 \pm 10\ meV$ are obtained in Fig. 9 (f). Moreover, with the determination of the trap DOS profile as explained by Khelifi et al. [52], we show that the density of the energy level observed with the addition of dopant (between 430 and 480 meV) increases by a factor 2 between 2% and 5% MR shown in Fig. S8 in supplementary material.

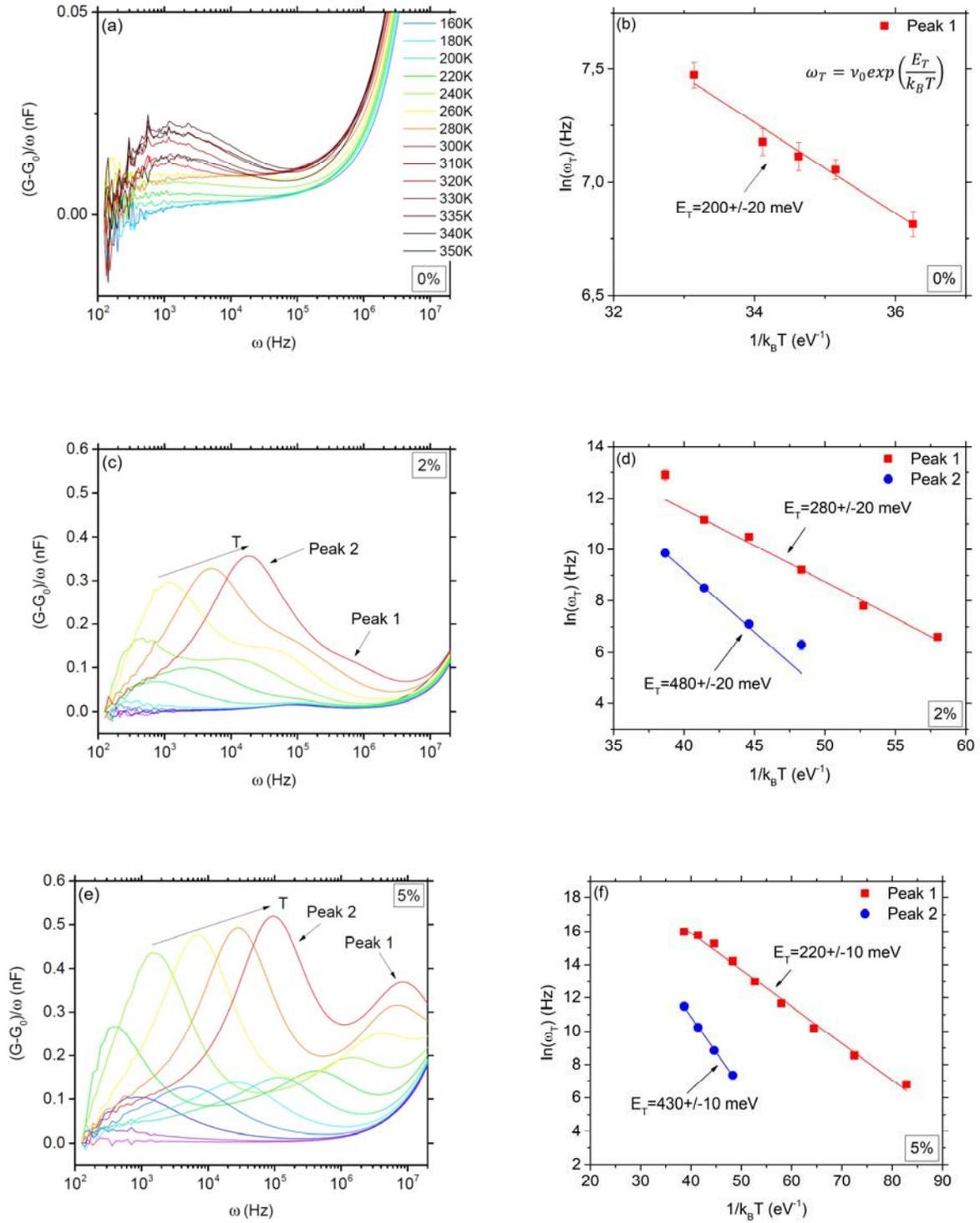

FIG. 9. $(G - G_0)/\omega$ versus angular frequency $\omega$ as a function of temperature at 0 V for MR=0% (a), 2% (c) and 5% (e). Arrhenius plot derived from the admittance spectroscopy measurements on the samples with doping concentrations of 0% (b), 2% (d) and 5% MR (f). The activation energies are extracted with a linear fit.

To determine whether the activation energies are given with respect to the polymer HOMO or LUMO, we can consider the electrodes work-function. To be probed by admittance spectroscopy, a trap state is necessarily crossed by the Fermi level for a measurement performed at 0 V. As a result, using admittance spectroscopy, we only probe the trap states situated between the electrodes work-function.[53] In the structure studied, the work-function of the anode in PEDOT:PSS has been measured by Kelvin probe at 4.9 eV, and the work-function of the cathode in aluminum is 4.3 eV.[54] Considering a HOMO at 5.15 eV and a LUMO at 3.35 eV below the vacuum level, activation energies between 200 and 480 meV are necessarily given with respect to the polymer HOMO. Fig. 10 (a) and (b) summarizes the energy levels probed by admittance spectroscopy with and without dopant.

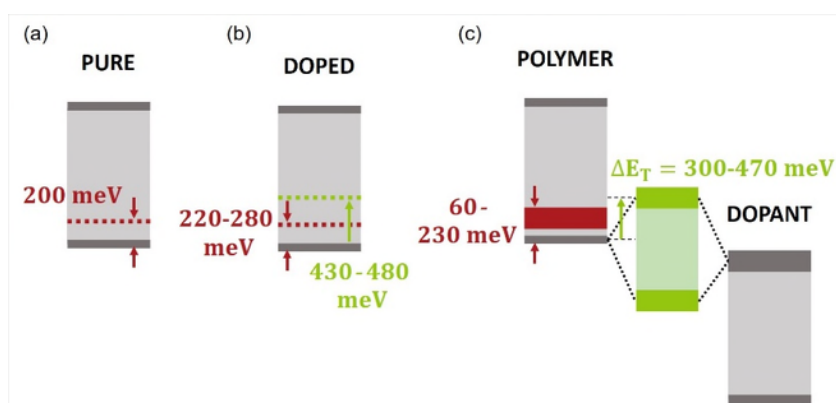

FIG. 10. Schematic band diagrams summarizing the trap states probed by admittance spectroscopy for pure (a) and doped (b) PBDTTT-c. Schematic of the hypothetical CTC formed between the PBDTTT-c HOMO and Mo(tfd-COCF$_3$)$_3$ LUMO with the energy levels obtained from UV-visible absorption spectroscopy considering a Hückel-like model.

The trap state identified in the pure and the doped polymer around 200-280 meV is situated approximately at the same position in the polymer bandgap. However, it is not straightforward to determine whether both trap states are due to the same origin as the position of the peak in the conductance spectrum depends on the capture cross-section and the carrier density, which evolves with doping. Considering that both levels correspond to the same trap distribution, an energy level intrinsic to PBDTTT-c might be situated around 200-280 meV above the polymer HOMO. With doping, an additional trap state is identified 430-480 meV above the polymer HOMO, which could correspond to the anti-bonding level of the CTC. Both energy levels probed by admittance spectroscopy are consistent with the hypothesis of CTC suggested from UV-visible absorption measurements and summarized in Fig. 10 (c).

**E. Effectiveness of p-doping**

Although our experiments suggest that the formation of a CTC is involved in the p-doping process of PBDTTT-c by Mo(tfd-COCF$_3$)$_3$, it is unclear how an acceptor level situated hundreds of meV above the polymer HOMO can lead to hole densities as high as $10^{18}\ cm^{-3}$ at a doping concentration of 1% MR (Fig. 3). This significant amount of energy required to p-dope the polymer weakens the hypothesis of CTC and prevents from correctly identifying the doping mechanism involved. Further work is necessary to understand what physical phenomenon could allow the transition from the polymer HOMO toward the complex antibonding state in case of CTC.

Arkhipov et al.[55] have shown that the addition of dopant leading to strong Coulomb interactions induces a DOS broadening with an increase of the deep states in the tail of the Gaussian distribution. A broadening of the tail states might lead to the overlap of the HOMO DOS and the distribution of antibonding states. Recent studies have also reported experimental results of DOS broadening with doping through UPS measurements. Pahner et al.[30] measured an increase of the tail states distribution upon addition of $C_{60}F_{36}$ in pentacene. This tail states evolution has also been observed by Zuo et al.[56] in $F_4$TCNQ doped P3HT. They developed a model showing good consistency with the data, highlighting the role of deep tail states in the evolution of the HOMO DOS. Lin et al.[57] have observed the evolution of the CuPc HOMO DOS when doped with Mo(tfd)$_3$. However, they showed that the addition of small amounts of dopant leads to the broadening of the main Gaussian distribution while the trap states in the tail are progressively filled with doping. Therefore, further studies are necessary to determine whether the evolution of the HOMO DOS might lead to an effective charge transfer from the polymer HOMO toward the potential antibonding state of the complex.

## IV. CONCLUSIONS

In this paper, we aimed to determine what doping mechanism is involved in Mo(tfd-COCF$_3$)$_3$ doped PBDTTT-c in order to further understanding the evolution of the hole density at high dopant concentration.

UV-visible absorption spectroscopy on pure and doped PBDTTT-c samples highlighted the formation of sub-bandgap absorption peaks with doping. The origin of these peaks might be due to the DCT mechanism with the formation of polarons in the polymer or to the formation of a CTC between the polymer HOMO and dopant LUMO. Considering the CTC hypothesis, the energy level associated with the complex were calculated using a Hückel-like model and suggested an antibonding state situated between 300 and 470 meV above the polymer HOMO.

To distinguish between DCT and CTC, we used photoluminescence spectroscopy showing the quenching of fluorescence upon dopant addition. The evolution with dopant concentration of the fluorescence intensity and of the excited state lifetime highlights the involvement of both static and dynamic quenching. This observation suggests that a CTC might be involved in the doping mechanism.

In order to probe the potential energy levels related to the CTC, we carried out admittance spectroscopy measurements on pure and doped PBDTTT-c. This experiment highlighted the presence of a trap state intrinsic to the polymer and situated around 200-280 meV above the polymer HOMO, and the formation of an additional trap state upon addition of dopant molecules situated between 430 and 480 meV above the polymer HOMO. The energy levels obtained are consistent with the CTC hypothesis formulated according to the UV-visible absorption measurements.

The combination of optical and electrical characterization techniques suggests that the formation of a CTC is involved in the p-doping of PBDTTT-c with $Mo(tfd-COCF_3)_3$. However, doubts can be raised regarding the CTC hypothesis as a significant amount of energy is required to allow the electron transfer from the polymer HOMO toward the anti-bonding state of the complex. Further studies need to be carried out to determine whether the DOS broadening upon dopant addition or other phenomena can explain the effective p-doping with CTC.

If the formation of a CTC is effectively involved in the doping process of $Mo(tfd-COCF_3)_3$ doped PBDTTT-c, it could explain the hole density sublinear increase at high doping concentration with a lower probability of dopant ionization. The p-doping ability of the dopant could be improved by reducing the gap of the CTC. This improvement could be achieved by reducing the intermolecular electronic coupling described by the resonance integral β.

## SUPPLEMENTARY MATERIAL

See supplementary material for supporting data and details.

## ACKNOWLEDGMENTS

We gratefully acknowledge the group of Prof. Seth Marder (Georgia Institute of Technology) for providing the dopants. We thank Joël Bleuse (CEA-INAC) for the TRPL measurements, Sylvie Lepilliet (IEMN-CNRS) for her help on the admittance spectroscopy set-up and measurements and Alexandre Pereira (CEA-LITEN) for the important comments on the absorption spectroscopy measurements. A.K.'s work at Princeton University was supported by a grant from the National Science Foundation (DMR-1506097).

**REFERENCES**

[1] S.M. Sze and K.K. Ng, *Physics of Semiconductor Devices* (John wiley & sons, 2006).

[2] K. Walzer, B. Maennig, M. Pfeiffer, and K. Leo, Chem. Rev. **107**, 1233 (2007).

[3] J. Huang, M. Pfeiffer, A. Werner, J. Blochwitz, K. Leo, and S. Liu, Appl. Phys. Lett. **80**, 139 (2002).

[4] G. He, O. Schneider, D. Qin, X. Zhou, M. Pfeiffer, and K. Leo, J. Appl. Phys. **95**, 5773 (2004).

[5] M. Pfeiffer, Sol. Energy Mater. Sol. Cells **63**, 83 (2000).

[6] B. Maennig, J. Drechsel, D. Gebeyehu, P. Simon, F. Kozlowski, A. Werner, F. Li, S. Grundmann, S. Sonntag, M. Koch, K. Leo, M. Pfeiffer, H. Hoppe, D. Meissner, N.S. Sariciftci, I. Riedel, V. Dyakonov, and J. Parisi, Appl. Phys. A Mater. Sci. Process. **79**, 1 (2004).

[7] S. Singh, S.K. Mohapatra, A. Sharma, C. Fuentes-Hernandez, S. Barlow, S.R. Marder, and B. Kippelen, Appl. Phys. Lett. **102**, (2013).

[8] T. Minari, T. Miyadera, K. Tsukagoshi, Y. Aoyagi, and H. Ito, Appl. Phys. Lett. **91**, 53508 (2007).

[9] S.P. Tiwari, W.J. Potscavage, T. Sajoto, S. Barlow, S.R. Marder, and B. Kippelen, Org. Electron. Physics, Mater. Appl. **11**, 860 (2010).

[10] H. Kleemann, C. Schuenemann, A.A. Zakhidov, M. Riede, B. Lüssem, and K. Leo, Org. Electron. **13**, 58 (2012).

[11] S.-J. Yoo and J.-J. Kim, Macromol. Rapid Commun. n/a (2015).

[12] S. Olthof, S. Mehraeen, S.K. Mohapatra, S. Barlow, V. Coropceanu, J.L. Brédas, S.R. Marder, and A. Kahn, Phys. Rev. Lett. **109**, 176601 (2012).

[13] M.L. Tietze, J. Benduhn, P. Pahner, B. Nell, M. Schwarze, H. Kleemann, M. Krammer, K. Zojer, K. Vandewal, and K. Leo, Nat. Commun. **9**, 1182 (2018).

[14] I. Salzmann and G. Heimel, J. Electron Spectros. Relat. Phenomena **204**, 1 (2015).

[15] B. Lüssem, M. Riede, and K. Leo, *Doping of Organic Semiconductors* (2013).

[16] P. Pingel and D. Neher, Phys. Rev. B - Condens. Matter Mater. Phys. **87**, 115209 (2013).

[17] H. Méndez, G. Heimel, S. Winkler, J. Frisch, A. Opitz, K. Sauer, B. Wegner, M. Oehzelt, C. Röthel, S. Duhm, D. Többens, N. Koch, and I. Salzmann, Nat. Commun. **6**, 8560 (2015).

[18] S. Yu, J. Frisch, A. Opitz, E. Cohen, M. Bendikov, N. Koch, and I. Salzmann, Appl. Phys. Lett. **106**, 203301 (2015).

[19] M.L. Tietze, L. Burtone, M. Riede, B. Lüssem, and K. Leo, Phys. Rev. B - Condens. Matter Mater. Phys. **86**, 1 (2012).

[20] C.K. Chan, W. Zhao, S. Barlow, S. Marder, and A. Kahn, Org. Electron. Physics, Mater. Appl. **9**, 575 (2008).

[21] C. Falkenberg, C. Uhrich, S. Olthof, B. Maennig, M.K. Riede, and K. Leo, J. Appl. Phys. **104**, 34506 (2008).

[22] G. Parthasarathy, C. Shen, A. Kahn, and S.R. Forrest, J. Appl. Phys. **89**, 4986 (2001).

[23] Y. Karpov, T. Erdmann, I. Raguzin, M. Al-Hussein, M. Binner, U. Lappan, M. Stamm, K.L. Gerasimov, T. Beryozkina, V. Bakulev, D. V Anokhin, D.A. Ivanov, F. Günther, S. Gemming, G. Seifert, B. Voit, R. Di Pietro, and A. Kiriy, Adv. Mater. **28**, 6003 (2016).

[24] I. Salzmann, G. Heimel, S. Duhm, M. Oehzelt, P. Pingel, B.M. George, A. Schnegg, K. Lips, R.P. Blum, A. Vollmer, and N. Koch, Phys. Rev. Lett. **108**, 1 (2012).

[25] L. Zhu, E. Kim, Y. Yi, and J.-L. Brédas, Chem. Mater. **23**, 5149 (2011).

[26] H. Méndez, G. Heimel, A. Opitz, K. Sauer, P. Barkowski, M. Oehzelt, J. Soeda, T. Okamoto, J. Takeya, J.B. Arlin, J.Y. Balandier, Y. Geerts, N. Koch, and I. Salzmann, Angew. Chemie - Int. Ed. **52**, 7751 (2013).

[27] G. Heimel, I. Salzmann, and N. Koch, **148**, 148 (2012).

[28] J. Euvrard, A. Revaux, P.-A. Bayle, M. Bardet, D. Vuillaume, and A. Kahn, Org. Electron. **53**, 135 (2018).

[29] A. Dai, Y. Zhou, A.L. Shu, S.K. Mohapatra, H. Wang, C. Fuentes-Hernandez, Y. Zhang, S. Barlow, Y.L. Loo, S.R. Marder, B. Kippelen, and A. Kahn, Adv. Funct. Mater. **24**, 2197 (2014).

[30] P. Pahner, H. Kleemann, L. Burtone, M.L. Tietze, J. Fischer, K. Leo, and B. Lüssem, Phys. Rev. B - Condens. Matter Mater. Phys. **88**, 195205 (2013).

[31] T. Kirchartz, W. Gong, S. a. Hawks, T. Agostinelli, R.C.I. MacKenzie, Y. Yang, and J. Nelson, J. Phys. Chem. C **116**, 7672 (2012).

[32] M. Pfeiffer, K. Leo, X. Zhou, J.S. Huang, M. Hofmann, A. Werner, and J. Blochwitz-Nimoth, Org. Electron. Physics, Mater. Appl. **4**, 89 (2003).

[33] S. Olthof, S. Mehraeen, S.K. Mohapatra, S. Barlow, V. Coropceanu, J.L. Brédas, S.R. Marder, and A. Kahn, Phys. Rev. Lett. **109**, 176601 (2012).

[34] Q. Bao, X. Liu, S. Braun, F. Gao, and M. Fahlman, Adv. Mater. Interfaces **2**, 1400403 (2015).

[35] F. Deschler, D. Riedel, A. Deák, B. Ecker, E. Von Hauff, and E. Da, Synth. Met. **199**, 381 (2015).

[36] M.L. Tietze, P. Pahner, K. Schmidt, K. Leo, and B. Lüssem, Adv. Funct. Mater. **25**, 2701 (2015).

[37] J.-P. Yang, W.-Q. Wang, F. Bussolotti, L.-W. Cheng, Y.-Q. Li, S. Kera, J.-X. Tang, X.-H. Zeng, and N. Ueno, Appl. Phys. Lett. **109**, 93302 (2016).

[38] I. Salzmann, G. Heimel, M. Oehzelt, S. Winkler, and N. Koch, Acc. Chem. Res. **49**, 370 (2016).

[39] S.K. Mohapatra, Y. Zhang, B. Sandhu, M.S. Fonari, T. V Timofeeva, S.R. Marder, and S. Barlow, Polyhedron **116**, 88 (2016).

[40] P. Lienhard, Caractérisation et Modélisation Du Vieillissement Des Photodiodes Organiques, Université Grenoble Alpes, 2016.

[41] J. Belasco, S.K. Mohapatra, Y. Zhang, S. Barlow, S.R. Marder, and A. Kahn, Appl. Phys. Lett. **105**, 2012 (2014).

[42] C.M. Cardona, W. Li, A.E. Kaifer, D. Stockdale, and G.C. Bazan, Adv. Mater. **23**, 2367 (2011).

[43] J. Sworakowski, Synth. Met. **235**, 125 (2018).

[44] A. Dai, Creating Highly Efficient Carrier Injection or Collection Contacts via Soft Contact Transfer Lamination of P-Doped Interlayers, Princeton University, 2015.

[45] S.R. Forrest, Chem. Rev. **97**, 1793 (1997).

[46] P.E. Shaw, A. Ruseckas, and I.D.W. Samuel, Adv. Mater. **20**, 3516 (2008).

[47] J. Yu, N.W. Song, J.D. Mcneill, and P.F. Barbara, Isr. J. Chem. **44**, 127 (2004).

[48] P. Tyagi, S. Tuli, and R. Srivastava, J. Chem. Phys. **142**, 54707 (2015).

[49] J.R. Lakowicz, *Principles of Fluorescence Spectroscopy* (Springer US, 1999).

[50] M.I.N. Zheng, F. Bai, F. Li, Y. Li, and D. Zhu, J. Appl. Polym. Sci. **70**, 599 (1998).

[51] J.A. Carr, M. Elshobaki, and S. Chaudhary, Appl. Phys. Lett. **107**, 203302 (2015).

[52] S. Khelifi, K. Decock, J. Lauwaert, H. Vrielinck, D. Spoltore, F. Piersimoni, J. Manca, A. Belghachi, and M. Burgelman, J. Appl. Phys. **110**, 94509 (2011).

[53] T. Walter, R. Herberholz, C. Müller, and H.W. Schock, J. Appl. Phys. **80**, 4411 (1996).

[54] H.B. Michaelson, J. Appl. Phys. **48**, 4729 (1977).

[55] V.I. Arkhipov, P. Heremans, E. V Emelianova, and H. Bässler, Phys. Rev. B - Condens. Matter Mater. Phys. **71**, 45214 (2005).

[56] G. Zuo, Phys. Rev. B **93**, 235203 (2016).

[57] X. Lin, G.E. Purdum, Y. Zhang, S. Barlow, S.R. Marder, Y.L. Loo, and A. Kahn, Chem. Mater. **28**, 2677 (2016).

# Toward a better understanding of the doping mechanism involved in Mo(tfd-COCF3)3 doped PBDTTT-c

J. Euvrard,[1] A. Revaux,[1,a)] S. S. Nobre,[2] A. Kahn,[3] D. Vuillaume[4]

[1]Univ. Grenoble Alpes, CEA-LITEN, Grenoble, 38000, France

[2]Univ. Grenoble Alpes, CEA, Liten, DTNM,SEN,LSIN, F-38000 Grenoble, France

[3]Dept. of Electrical Engineering, Princeton University, Princeton, NJ, 08544, USA

[4]IEMN, CNRS, Univ. Lille, Villeneuve d'Ascq, 59652, France

**SUPPLEMENTARY MATERIAL**

## A. Excited state lifetime extraction

To extract the excited state lifetime, we fit the TRPL spectra with a double exponential function as follows:

$$y = y_0 + A_1 \exp(-x/\tau_1) + A_2 \exp(-x/\tau_2). \tag{1}$$

As intensity values evolve over several orders of magnitude, we weight the data using standard deviation. We also avoid to fit the tail of the signal where the signal-to-noise ratio is low. The extracted lifetimes are given in Table S1.

Table S1. Lifetimes extracted from the TRPL spectra fitting with a double exponential.

| MR (%) | Extracted $\tau_1$ (ps) | Extracted $\tau_2$ (ps) | Corrected $\tau_2$ (ps) |
|---|---|---|---|
| 0 | 419 | 850 | 780 ± 10 |
| 0.5 | 320 | 750 | 680 ± 10 |
| 1 | 280 | 660 | 570 ± 10 |
| 2 | 240 | 550 | 440 ± 20 |
| 4 | 230 | 480 | 350 ± 20 |

The extracted values of $\tau_1$ and $\tau_2$ are short (below 1 ns) and close to the time response of the APD (between 300 and 350 ps). The values of $\tau_1$ extracted during the first nanosecond of decay and lower than 350 ps for doped polymer are probably related to the APD response. The value $\tau_2$ is related to the excited state lifetime of pure and doped polymer, and needs to be deconvoluted from the APD response. If we consider the APD response time $\tau_{APD}$ and the PBDTTT-c excited state lifetime $\tau_{PBDTTT-c}$ as two independent variables, the measured time constant $\tau$ is given by:

$$\tau^2 = \tau_{APD}^2 + \tau_{PBDTTT-c}^2. \tag{2}$$

Using this relation, the PBDTTT-c excited state lifetime $\tau_{PBDTTT-c}$ is extracted for each dopant concentration and given in Table S1.

## B. Supplementary figures

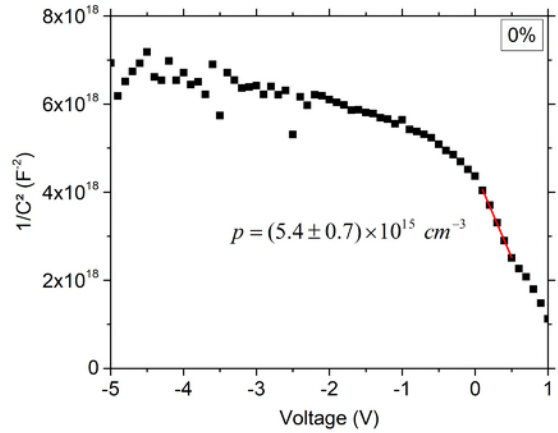

FIG. S1. Mott-Schottky plot for pure PBDTTT-c. The hole density is indicated.

For the pure polymer we obtain a concentration around $5.10^{15}$ $cm^{-3}$ using the Mott-Schottky analysis as shown in Fig. S1. However, Kirchartz et al.[1] have highlighted the limits of the Mott-Schottky extraction technique at low doping densities. They determined that, for a diode with a Schottky contact and a layer thickness around 300 nm, the Mott-Schottky extraction technique is not appropriate for hole concentrations below $4.10^{15}$ $cm^{-3}$ because the depletion approximation is not valid. With a hole density around $5.10^{15}$ $cm^{-3}$, it is difficult to consider the value extracted using the Mott-Schottky relation as reliable. Therefore, we preferred to start our analysis of hole density at a doping concentration of 0.5% where our Mott-Schottky extraction technique is more reliable.

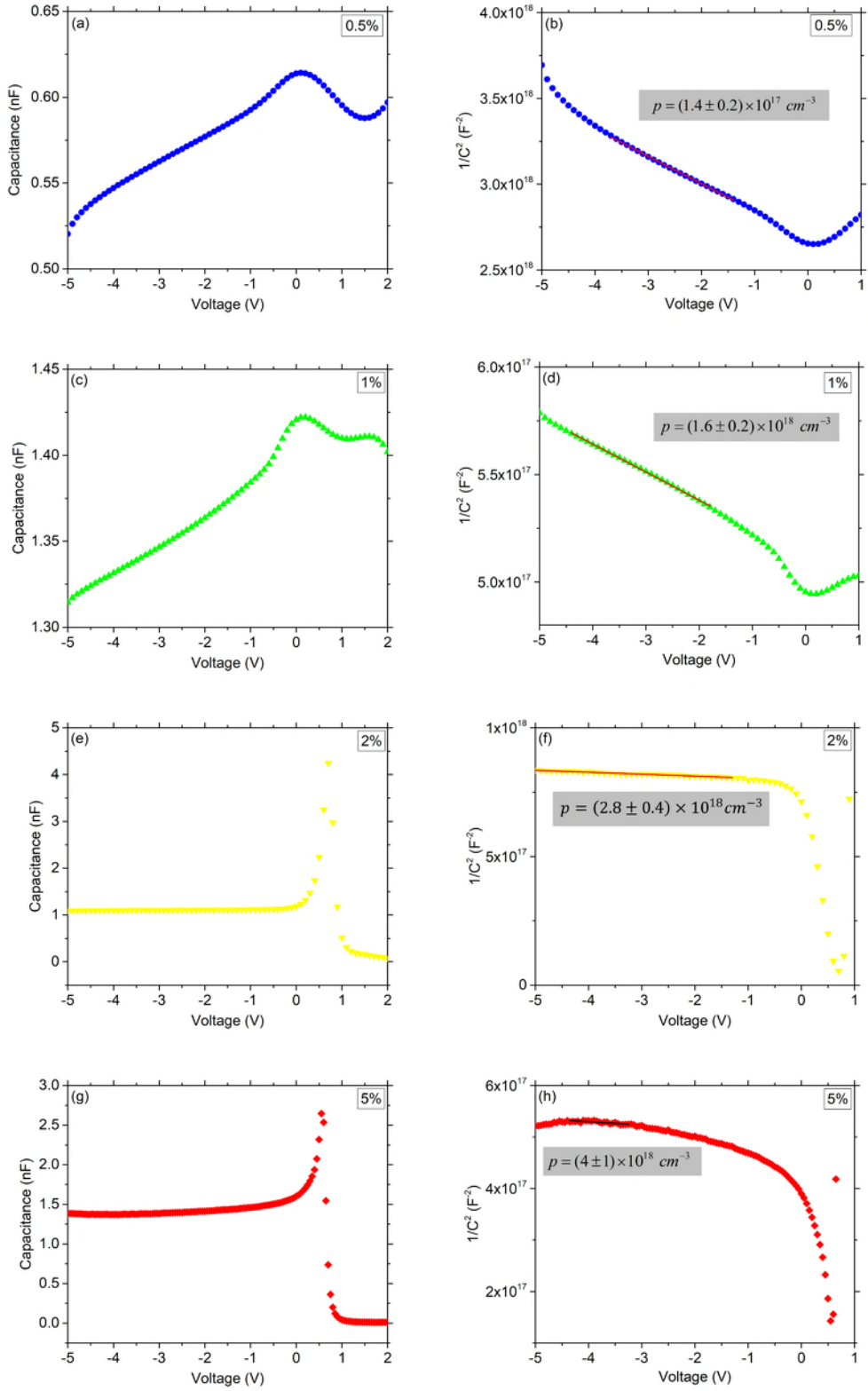

FIG. S2. C(V) characteristic (a), (c), (e), (g) and Mott-Schottky plot (b), (d), (f), (h) for doped PBDTTT-c at 0.5%, 1%, 2% and 5% MR.

The C(V) characteristics for 0.5%, 1%, 2% and 5% MR are shown in Fig. S2 (a), (c), (e) and (g) respectively. To extract the hole density, we use the Mott-Schottky representation given in Fig. S2 (b), (d), (f) and (h). For a homogeneous doping, the hole density p can be extracted from the linear dependency of $1/C^2$ in reverse bias. For the doped samples, no linear behavior can be obtained over the full bias range as the doping is not homogeneous in the device. At low reverse bias, the pure PBDTTT-c layer at the aluminum electrode layer is depleted. The hole density extracted in this regime corresponds to the pure layer. At higher reverse bias, the pure layer is totally depleted and the depletion reaches the doped central layer of polymer. Therefore, it is important to determine at what biases the Mott-Schottky analysis needs to be carried out.

According to Khelifi et al.[2], the hole density profile in the semiconductor layer can be obtained from the Mott-Schottky law generalized for inhomogeneous doping:

$$\frac{d(1/C_j^2)}{dV} = -\frac{2}{qA^2\varepsilon_0\varepsilon_r p(w)} \tag{3}$$

where q is the elementary charge, A the diode area, $\varepsilon_0$ and $\varepsilon_r$ the vacuum and relative permittivities, w the depth in the semiconductor layer given by the capacitance at a given voltage $C_j$:

$$C_j = \frac{A\varepsilon_0\varepsilon_r}{w}. \tag{4}$$

As an example, the profile of the hole density is given for the 1% MR sample at 300 K in Fig. S3. A plateau is reached above 87.5 nm (from the aluminum electrode). Below 87.5 nm, the hole density of the pure polymer should be measured as only the pure layer is depleted. This threshold corresponds to an applied bias of -1 V. When the applied bias is lower than -1 V, the depletion reaches the doped layer leading to the desired value of p. This graph illustrates the origin of the two slopes observed in the Mott-Schottky plots for the samples with a doped layer. Therefore, only the second slope is taken into account for the Mott-Schottky analysis for each doping concentration as shown in Fig. S2 (b), (d), (f) and (h).

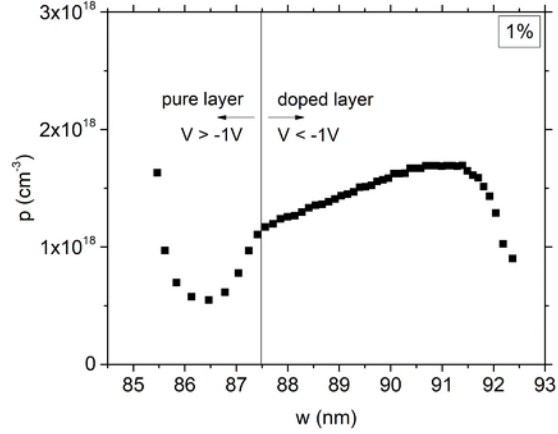

FIG. S3. Hole density profile in the semiconductor layer for the 1% MR sample measured at ambient temperature.

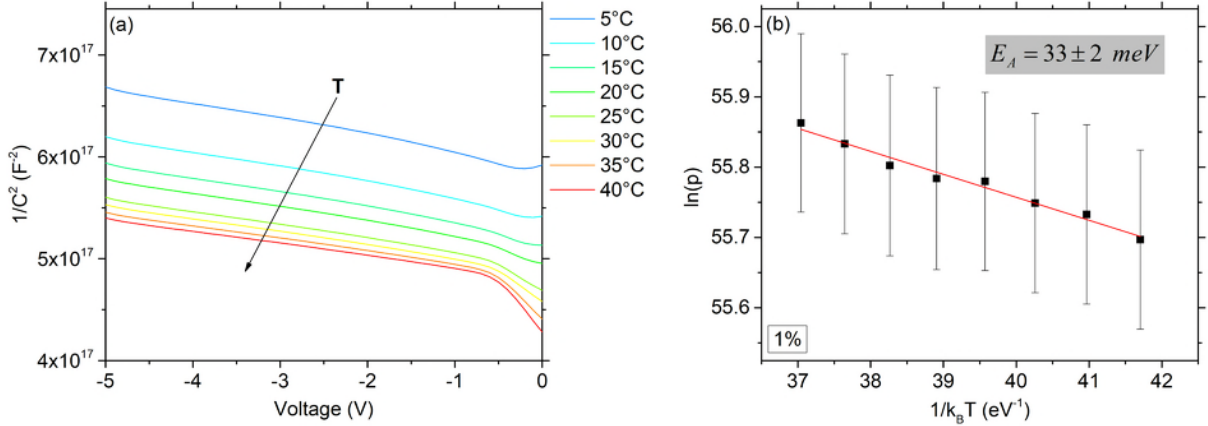

FIG. S4. Mott-Schottky representation of capacitance measurements from 5°C to 40°C for 1% MR (a) and corresponding Arrhenius plot of the doping activation with temperature (b).

We have carried out temperature-dependent C(V) measurements in order to extract the doping activation energy. Unfortunately, our extraction is limited by the significant uncertainty of the hole density through Mott-Schottky analysis and the limited temperature range permitted for this analysis. Below 0°C, capacitance measurements with respect to the applied bias exhibit a very low signal. Moreover, we have chosen to limit temperature measurements of these devices at 40°C to avoid potential dopant diffusion issues in the pure layers. The Mott-Schottky plots for the sample doped at 1% MR are given in Fig. S4 (a) for temperatures varying from 5°C to 40°C. We can observe a slight decrease of the slope in the reverse bias regime with temperature. The corresponding Arrhenius plot is shown in Fig. S4 (b). A linear fit of these data leads to a thermal dopant activation energy of 33 ± 2 meV. However, the thermal dopant activation energy extraction is limited by the significant uncertainty obtained for the hole density and the short temperature range studied.

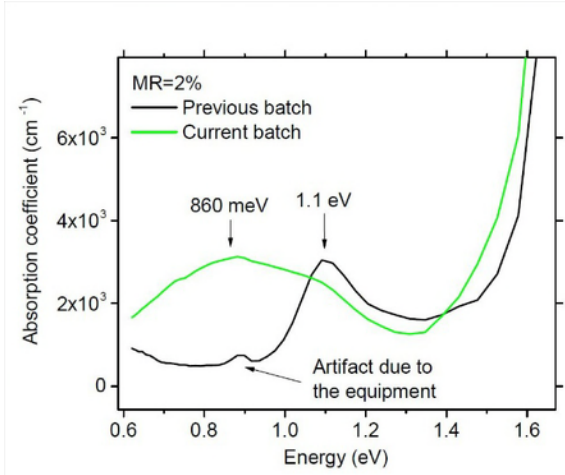

FIG. S5. Absorption coefficient spectra for two batches of PBDTTT-c doped at 2% MR with Mo(tfd-COCF$_3$)$_3$.

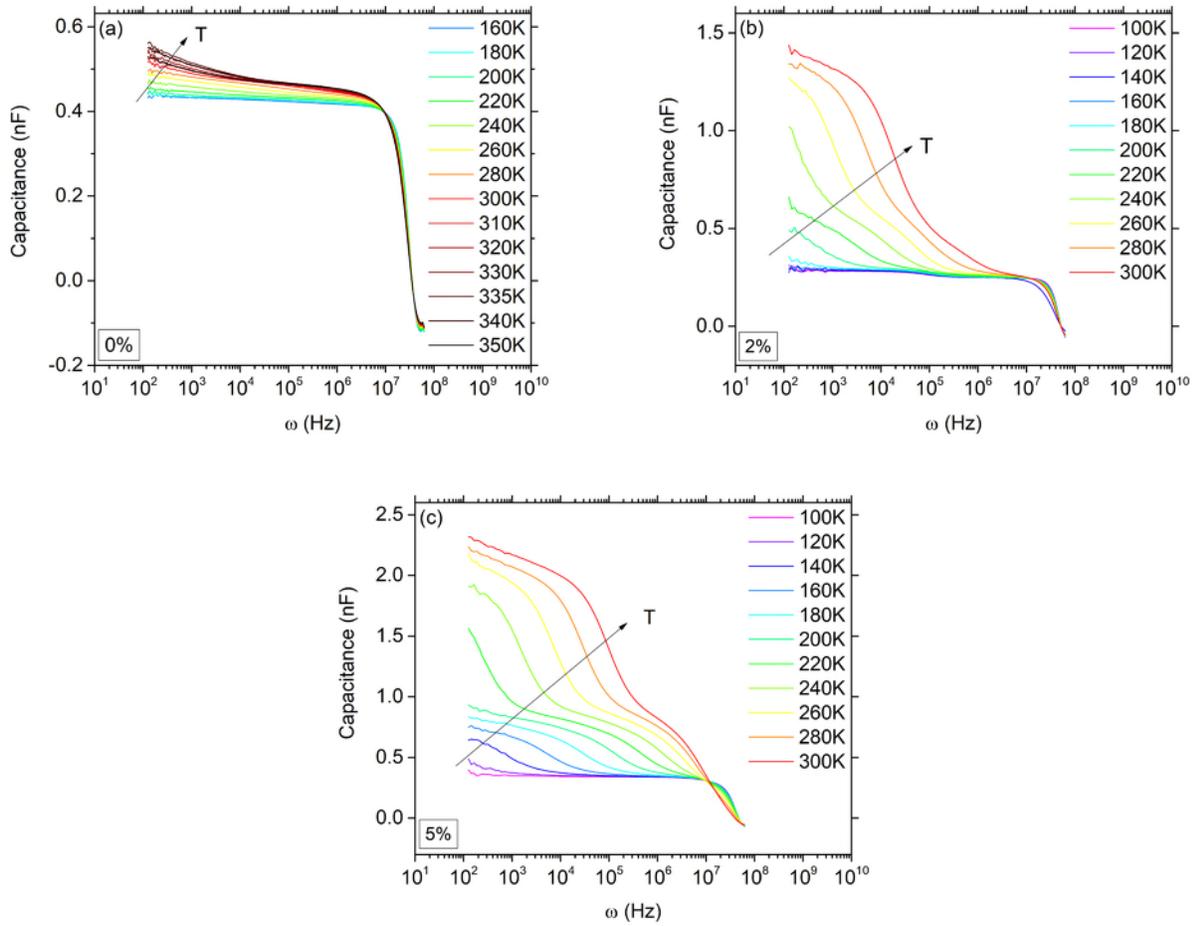

Fig. S6. Capacitance vs angular frequency as a function of temperature at 0 V for MR=0% (a), 2% (b) and 5% (c) respectively.

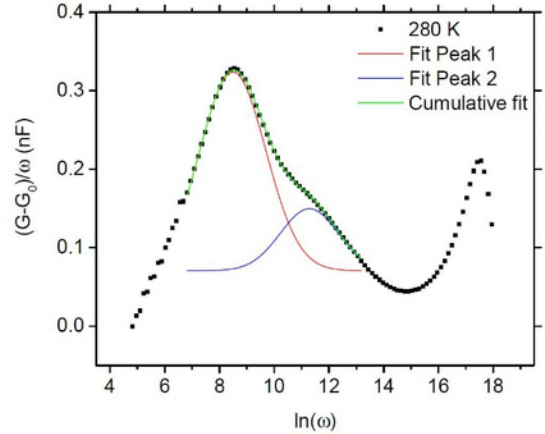

FIG. S7. $(G - G_0)/\omega$ vs angular frequency ω at 280 K for the 2% MR sample and Gaussian fitting for peaks 1 and 2.

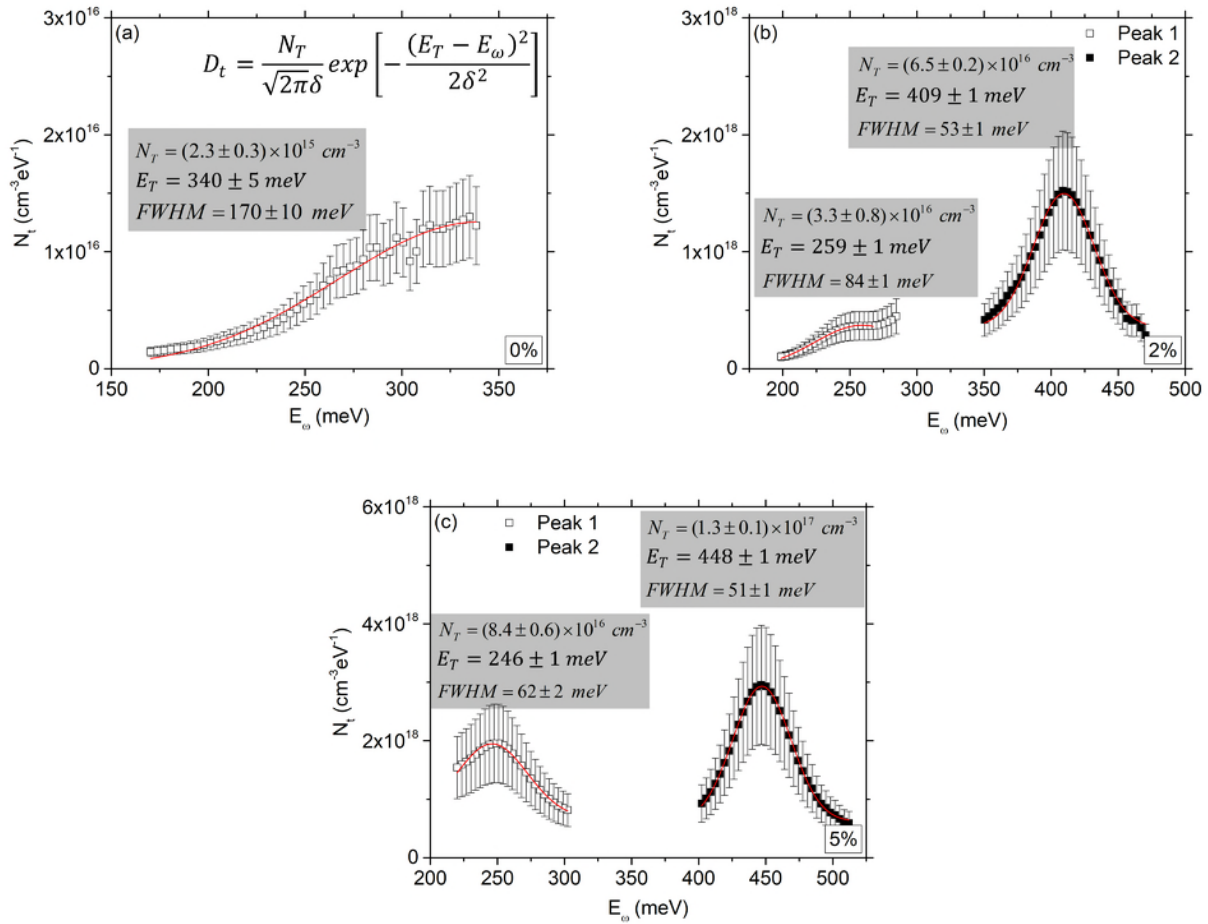

FIG. S8. DOS profiles at 300 K for MR=0% (a), 2% (b) and 5% (c). The fitting parameters are indicated.

**SUPPLEMENTARY MATERIAL REFERENCES**

[1] T. Kirchartz, W. Gong, S. a. Hawks, T. Agostinelli, R.C.I. MacKenzie, Y. Yang, and J. Nelson, J. Phys. Chem. C 116, 7672 (2012).

[2] S. Khelifi, K. Decock, J. Lauwaert, H. Vrielinck, D. Spoltore, F. Piersimoni, J. Manca, A. Belghachi, and M. Burgelman, J. Appl. Phys. 110, 94509 (2011).

[2] S. Khelifi, K. Decock, J. Lauwaert, H. Vrielinck, D. Spoltore, F. Piersimoni, J. Manca, A. Belghachi, and M. Burgelman, J. Appl. Phys. 110, 94509 (2011).